\documentclass{article}

\usepackage{PRIMEarxiv}

\usepackage[utf8]{inputenc} 
\usepackage[T1]{fontenc}    
\usepackage{hyperref}       
\usepackage{url}            
\usepackage{booktabs}       
\usepackage{amsfonts}       
\usepackage{nicefrac}       
\usepackage{microtype}      
\usepackage{lipsum}
\usepackage{fancyhdr}       
\usepackage{graphicx}       
\usepackage{algorithmic}
\usepackage{algorithm}
\usepackage{subcaption}
\usepackage{multirow}
\usepackage{multicol}
\usepackage{lscape}
\graphicspath{{media/}}     

\title{Dataloader Parameter Tuner: An Automated Dataloader Parameter Tuner for Deep Learning Models
}

\author{JooYoung Park, DoangJoo Synn\\
Korea University \\ 
Seoul\\
Republic of Korea\\
\texttt{\{nehalem, alansynn\}@korea.ac.kr}\\
\And
XinYu Piao\\
Korea University\\
Seoul\\
Republic of Korea\\
\texttt{xypiao97@korea.ac.kr}\\
\And
Jong-Kook Kim\\
Korea University\\
Seoul\\
Republic of Korea\\
\texttt{jongkook@korea.ac.kr}\\
}

\begin{document}
\maketitle

\begin{abstract}
\label{section:abstract}

Deep learning has recently become one of the most compute/data-intensive methods and is widely used in many research areas and businesses.
One of the critical challenges of deep learning is that it has many parameters that can be adjusted, and the optimal value may need to be determined for faster operation and high accuracy. The focus of this paper is the adjustable parameters of the dataloader. The dataloader in a system mainly groups the data appropriately and loads it to the main memory for the deep learning model to use. 
We introduce an automated framework called Dataloader Parameter Tuner (DPT) that determines the optimal value for the parameters required for the dataloader.
This framework discovers the optimal values for the number of dataloader's subprocesses (i.e., worker) and prefetch factor through grid search to accelerate the data transfer for machine learning systems.
\end{abstract}

\section{Introduction}
Recently, data collected by many enterprises are increasing in capacity, resolution, and variety.
This is due to the increase in the use of mobile devices and the Internet. 
To process this so called ‘Big Data’, the need for Deeper Neural Networks (DNN) is also increasing.
Conversely, bigger datasets are needed to successfully train more complex DNN \cite{is-your-dataset-big-enough}, both leading to need for more computing power and more memory. 
To fulfill the demand for more computing power, modern computing architectures use multi-CPUs and GPUs \cite{multigpu-cuda-opengl, multigpu-strategy}, which exhibits the characteristics of a distributed computing system in a single node or system. 
Therefore, various learning techniques such as data parallelism \cite{ddp:pipebw, ddp:soap, ddp:ssp}, model parallelism \cite{modelparallelism-adam, modelparallelism-largescale}, and pipeline parallelism \cite{gpipe, pipedream, mbs} are introduced for effecient use of such systems. 

Modern deep learning framework tries to utilize dataloader as much as possible by using multi-threading (e.g. Tensorflow) \cite{tensorflow-parallel} and multi-processing (e.g. PyTorch) \cite{pytorch-parallel}. 
As default, the main dataloader of a system mostly loads, reprocesses, and shuffles data and passes it to the targeted DNN. 
To take advantage of parallelism, the main dataloader process spawns multiple sub process, also known as worker or thread. 
And the number of these sub processes can be adjusted by using arguments.

Recent studies for dataloaders are mostly application specific implementations like dataloader for language related applications \cite{dataloader-language}, or graph related applications \cite{dataset-graph}.
Because of diverse computing environments and various DNN models, tuning common parameters such as workers/threads and prefetch factors may be the solution to overall performance boost. Deep learning frameworks such as PyTorch have default parameter values for the dataloader which are half of CPU cores for the number workers and 2 for the prefetch factor. Due to varying computing environment, these parameter values are often not the optimal. Number of CPU cores, CPU performance, system memory, number of GPU and its performance, and even GPU Memory size and system I/O performance makes it difficult to determine the optimal parameter values. 
This paper proposes the Dataloader Parameter Tuner (DPT) that determines the optimal number of workers and prefetch factor for a particular system. Thus maximizing the effectiveness of the dataloader.

This paper is organized as follows. The concept of dataloader and the parameter to tune are described in Section \ref{section:dataloader}. 
Section \ref{section:dpt} introduces the DPT method and terminologies used throughout this paper. 
Experimental results are depicted in Section \ref{section:experiments}.
Finally, Section \ref{section:conclusion} summarizes the research.

\section{Dataloader}
\label{section:dataloader}
\subsection{Overview}
A dataloader process usually consists of four steps. First two steps are dataloading and transform. Dataloader first calls a dataset instance. The dataset instance reads the data  and label from the storage, and then transforms (e.g., padding, tensorize, collate function) them to a suitable form for future processes. Third is shuffling and batching according to the arguments that the dataloader received. Fourth is prefetching. Prefetching loads \(n\) samples in advance such that communication latency can be hidden.
%

%



%

\subsection{Worker}
In Pytorch, the dataloader has a parameter called \texttt{num worker}, which indicates the number of workers to be spawned. Dataloader spawns a number of worker processes, where each worker receives a series of arguments which include the location of dataset, collate function, and values to initialize the worker. This means data is accessed through disk I/O, and the transformation of dataset is executed within the worker process. These workers prepare the data in parallel.





\section{Dataloader Parameter Tuner}
\label{section:dpt}

%

\subsection{Overview}
\begin{figure}[h]
    \centering
    \includegraphics[scale=0.6]{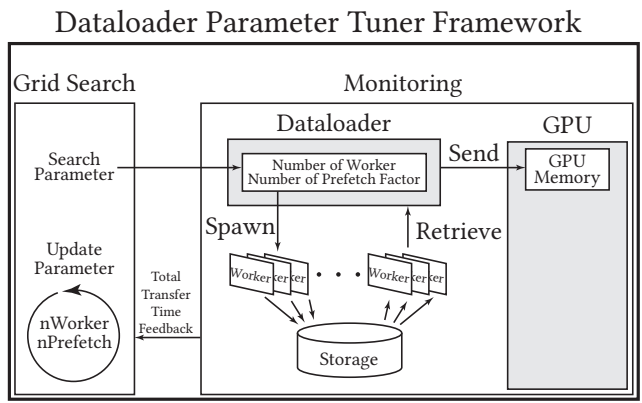}
    \caption{Overview of DPT}
    \label{fig:overview_dpt}
\end{figure}

Dataloader Parameter Tuner (DPT) is a framework that finds the optimal parameters for dataloader using grid search (Figure \ref{fig:overview_dpt}).
A grid search is a methodology to find a optimal value by substituting all combinations of the candidate hyper parameters.
Using grid search allows DPT to take variables that are difficult to parameterize into account such as dataset properties and hardware dependency.
And parameters drawn from DPT may be reused on the same machine upon loading data sets that have similar characteristics.

\subsection{Procedure}
%
In Algorithm \ref{alg_cap1}, DPT tries to determine optimal values for two dataloader parameters. The first one is the  \texttt{nWorker} which is the number of workers spawned by the dataloder, the other is the \texttt{nPrefetch}, which is the number of batches to be processed ahead of time. 
In the beginning of DPT, initialized values of the three variables \texttt{N} which is the number of available CPU cores, \texttt{G} which is the number of GPUs, and \texttt{P} which is the predefined maximum prefetch factor are needed.
%
Then, DPT iteratively determines the optimal combination of \texttt{nWorker} and \texttt{nPrefetch} using the transfer time that has occurred between main memory and main storage.

DPT starts from a state that has \texttt{G} workers spawned, and it is increased by \texttt{G} at every iteration.
The reason for starting and increasing the number of workers by \texttt{G} is because if \texttt{nWorker} is not a multiple of \texttt{G}, uneven number of workers are allocated to GPUs, causing execution speed difference, which leads to performance bottleneck.
%
For every \texttt{nWorker} step, DPT will calculate the transfer time using \texttt{nPrefetch} up to \texttt{P}. However, when memory overflow occurs the inner while loop will stop and go on to the next \texttt{nWorker} step.
%
%
%
%
%
%
\begin{algorithm}[h]
    \centering
    \begin{algorithmic}[1]
    \caption{Dataloader Parameter Tuner (DPT)}
    \label{alg_cap1}
    \REQUIRE Number of CPU cores $N$, Number of available GPUs $G$, Predefined prefetch factor $P$
    \ENSURE Number of workers $nWorker$, Number of prefetch factor $nPrefetch$

    \STATE $nWorker, nPrefetch \gets 0$ 
    \STATE $optimal{\_}time \gets \infty$
    \STATE $i \gets 0$
        \WHILE{$i < N$}
            \STATE$ i \gets i + G$
            \STATE $j \gets 0$
            \WHILE{$j < P$}
            \STATE Initialize Main Memory to measure Dataloader Transfer Time
            \IF{Memory Overflow occur}
                \STATE \textbf{break}
            \ELSE
                \STATE $total{\_}time \gets$ Measure Dataloader Transfer Time using $i, j$ arguments
            \ENDIF
            \IF{$total{\_}time < optimal{\_}time$}
                \STATE $optimal{\_}time \gets total{\_}time$
                \STATE $nWorker \gets i$
                \STATE $nPrefetch \gets j$
            \ENDIF
            \STATE$ j \gets j + 1$
            \ENDWHILE
        \ENDWHILE

    \end{algorithmic}
\end{algorithm}
\section{Experiments}
\label{section:experiments}

\subsection{Experimental Setup}
All experiments are run on a system consisting of Ubuntu 16.04 LTS operating system with NVIDIA 3080 Ti off-the-shelf GPU with 12 GB of graphic memory and 6 core 12 thread Intel Core i7-8700K 3.7GHz CPU with 64GB main memory. 
%
The environment used CUDA 11.3, and Pytorch 1.10.0. All experiments are performed with Pinned Memory option turned on.
%
%

CIFAR-10 \cite{cifar-10} and COCO 2017 Unlabeled images \cite{coco} dataset are used for the experiments.
CIFAR-10 is an image classification dataset with approximately 60K 32\(\times\)32 images in 10 classes.
COCO 2017 Unlabeled images dataset is also an image classification dataset with approximately 123K unlabeled images, and varies in image resolutions, and total dataset size is around 19GB. 
%
%
%
\begin{figure}[t]
        \begin{subfigure}[h]{0.5\linewidth}
            \includegraphics[scale=0.29]{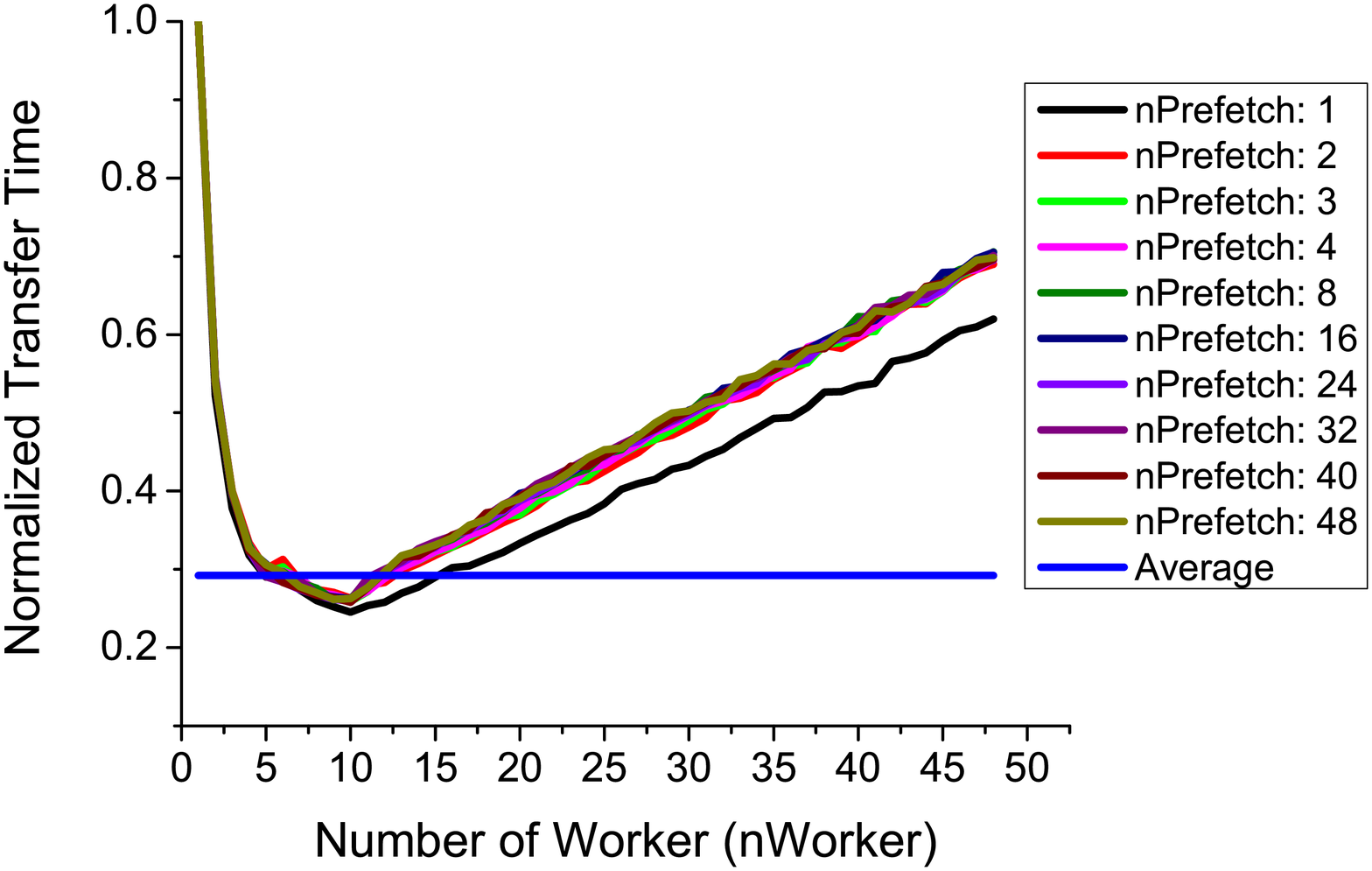}
            \caption{Normalized Transfer Time of Various Prefetch Factors While Increasing the Number of Workers}
            \label{fig:grid-worker}
        \end{subfigure}
        \begin{subfigure}[h]{0.5\linewidth}
            \includegraphics[scale=0.29]{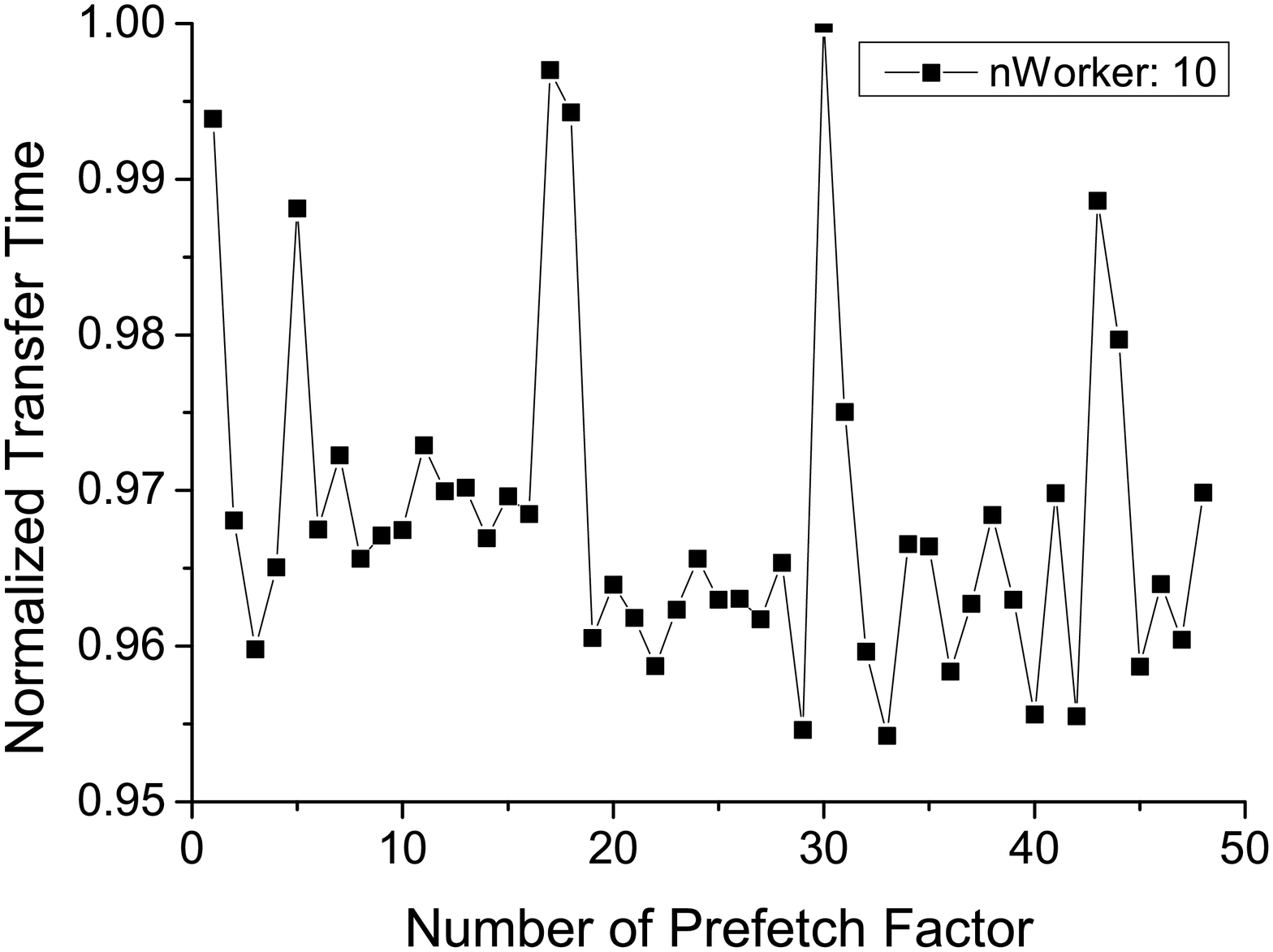}
            \caption{Result of Different Prefetch Factors for 10 Workers}
            \label{fig:grid-prefetch}
        \end{subfigure}
    \caption{Normalized Transfer Time}
    \label{fig:grid}
\end{figure}
\begin{figure}[h!]
    \centering
    \includegraphics[scale=0.275]{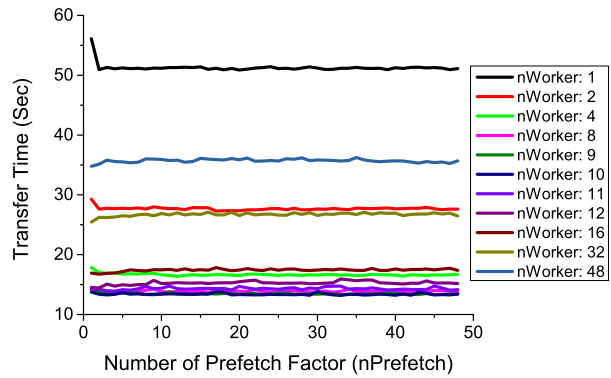}
    \caption{Transfer Time of Various Number of Workers While Increasing the Prefetch Factor}
    \label{fig:raw}
\end{figure}

\begin{figure*}[t]
    \centering
    \includegraphics[scale=0.6]{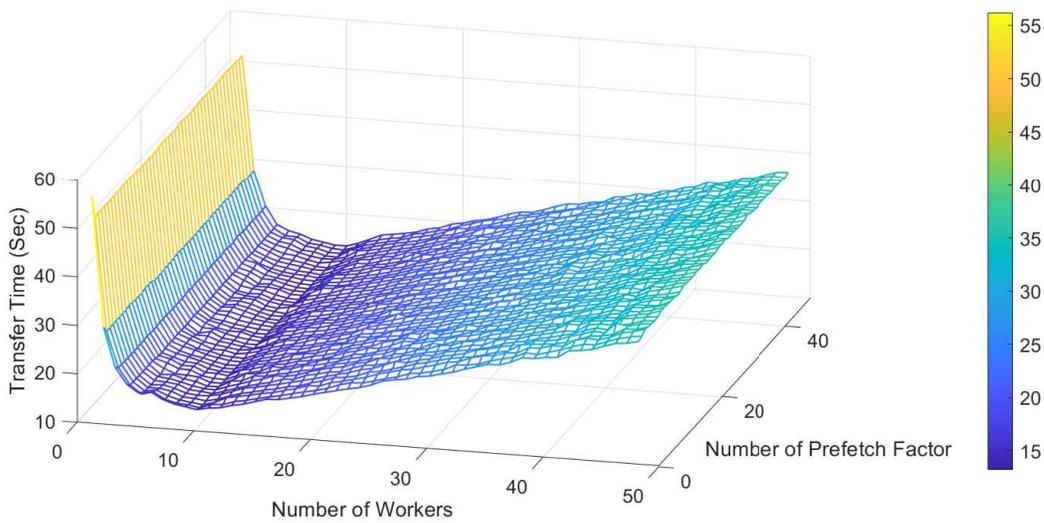}
    \caption{3D Representation of the Grid Search Using CIFAR-10}
    \label{fig:3D}
\end{figure*}
\subsection{CIFAR-10 Dataset}
\label{sec:grid_search}

\begin{landscape}
\begin{table}[b]
    \begin{subtable}[h]{0.45\linewidth}
    \centering
    \caption{Optimal Number of Workers}
    \label{tab:optimal-worker}
    \begin{tabular}{cc|cccc}
    \toprule
    \multirow{2}{*}{Batch} & \multirow{2}{*}{Epoch} & \multicolumn{4}{c}{Optimal Number of Workers}                 \\ 
    \cline{3-6}            &                        & 80\(\times\)80  & 160\(\times\)160 & 320\(\times\)320 & 640\(\times\)640\\
    \midrule
    \multirow{2}{*}{16}    & 1st                    & 10            & 10            & 5             & 6             \\
                           & After 2nd              & 9             & 9             & 9             & 7             \\ \hline
    \multirow{2}{*}{32}    & 1st                    & 10            & 10            & 5             & 5             \\
                           & After 2nd              & 10            & 9             & 9             & 8             \\ \hline
    \multirow{2}{*}{64}    & 1st                    & 10            & 10            & 5             & 6             \\
                           & After 2nd              & 10            & 10            & 9             & 7             \\ \hline
    \multirow{2}{*}{128}   & 1st                    & 10            & 10            & 5             & 5             \\
                           & After 2nd              & 10            & 10            & 8             & 8             \\ \hline
    \multirow{2}{*}{256}   & 1st                    & 10            & 10            & 5             & 6             \\
                           & After 2nd              & 10            & 10            & 10            & 7             \\ \hline
    \multirow{2}{*}{512}   & 1st                    & 10            & 10            & 6             & 6             \\
                           & After 2nd              & 10            & 10            & 9             & 5             \\ \hline
    \multirow{2}{*}{1024}  & 1st                    & 10            & 10            & 5             & N/A           \\
                           & After 2nd              & 10            & 10            & 10            & N/A           \\
    \bottomrule
    \end{tabular}
\end{subtable}
\hfill
    \begin{subtable}[h]{0.45\linewidth}
    \centering
    \caption{Transfer Time in Seconds}
    \label{tab:transfer_time}
    \begin{tabular}{cc|cccc}
    \toprule
    \multirow{2}{*}{Batch} & \multirow{2}{*}{Epoch} & \multicolumn{4}{c}{Transfer time (Sec)}                 \\ 
    \cline{3-6}            &                        & 80\(\times\)80  & 160\(\times\)160 & 320\(\times\)320 & 640\(\times\)640\\
    \midrule
    \multirow{2}{*}{16}    & 1st                    & 409.61        & 500.96        & 772.46        & 1290.03       \\
                           & After 2nd              & 8.67          & 23.57         & 89.48         & 346.94        \\ \hline
    \multirow{2}{*}{32}    & 1st                    & 411.44        & 499.42        & 768.12        & 1279.60       \\
                           & After 2nd              & 7.75          & 22.70         & 88.94         & 346.82        \\ \hline
    \multirow{2}{*}{64}    & 1st                    & 408.39        & 501.72        & 771.40        & 1288.92       \\
                           & After 2nd              & 7.50          & 22.22         & 87.95         & 344.91        \\ \hline
    \multirow{2}{*}{128}   & 1st                    & 404.66        & 499.23        & 771.04        & 1274.99       \\
                           & After 2nd              & 7.40          & 21.55         & 87.51         & 343.89        \\ \hline
    \multirow{2}{*}{256}   & 1st                    & 403.58        & 502.99        & 772.38        & 1286.86       \\
                           & After 2nd              & 7.26          & 21.55         & 87.08         & 345.72        \\ \hline
    \multirow{2}{*}{512}   & 1st                    & 395.74        & 507.89        & 813.92        & 1305.27       \\
                           & After 2nd              & 7.27          & 21.93         & 87.68         & 598.26        \\ \hline
    \multirow{2}{*}{1024}  & 1st                    & 400.72        & 495.93        & 779.36        & N/A           \\
                           & After 2nd              & 4.31          & 22.23         & 87.95         & N/A           \\
    \bottomrule
    \end{tabular}
\end{subtable}
\hfill
    \begin{subtable}[h]{0.45\linewidth}
    \centering
    \caption{Time Gain}
    \label{tab:overall time gain}
    \begin{tabular}{cc|cccc}
    \toprule
    \multirow{2}{*}{Batch} & \multirow{2}{*}{Epoch} & \multicolumn{4}{c}{DPT Time Reduction (\%)}                 \\ 
    \cline{3-6}            &                        & 80\(\times\)80  & 160\(\times\)160 & 320\(\times\)320 & 640\(\times\)640\\
    \midrule
    \multirow{2}{*}{16}    & 1st                    &-14.55 	&-12.69 	&-4.17 	    &0.00     \\
                           & After 2nd              &-24.92 	&-13.79 	&-5.70 	    &-2.28                            \\ \hline
    \multirow{2}{*}{32}    & 1st                    &-14.42 	&-13.65 	&-4.28 	    &-1.35     \\
                           & After 2nd              &-27.25 	&-11.06 	&-3.83 	    &-1.65                            \\ \hline
    \multirow{2}{*}{64}    & 1st                    &-16.04 	&-13.82 	&-4.28 	    &0.00     \\
                           & After 2nd              &-20.39 	&-7.67 	    &-2.88 	    &-1.12                            \\ \hline
    \multirow{2}{*}{128}   & 1st                    &-16.32 	&-12.62 	&-5.07 	    &-1.49     \\
                           & After 2nd              &-17.14 	&-6.89 	    &-2.71 	    &-1.82                            \\ \hline
    \multirow{2}{*}{256}   & 1st                    &-15.74 	&-12.88 	&-4.64 	    &0.00     \\
                           & After 2nd              &-18.28 	&-5.58 	    &-4.51 	    &-1.56                            \\ \hline
    \multirow{2}{*}{512}   & 1st                    &-16.81 	&-11.48 	&0.00 	    &0.00     \\
                           & After 2nd              &-17.49 	&-8.40 	    &-3.84 	    &-12.22                            \\ \hline
    \multirow{2}{*}{1024}  & 1st                    &-16.28 	&-19.53 	&-29.08     & N/A           \\
                           & After 2nd              &-21.85 	&-11.71 	&-4.41      & N/A           \\
    \bottomrule
    \end{tabular}
\end{subtable}
    \hfill
    \begin{subtable}[h]{0.45\linewidth}
    \centering
    \caption{Speed Up}
    \label{tab:DPT speedup}
    \begin{tabular}{cc|cccc}
    \toprule
    \multirow{2}{*}{Batch} & \multirow{2}{*}{Epoch} & \multicolumn{4}{c}{Speed Up}                 \\ 
    \cline{3-6}            &                        & 80\(\times\)80  & 160\(\times\)160 & 320\(\times\)320 & 640\(\times\)640\\
    \midrule
    \multirow{2}{*}{16}    & 1st                    &1.17\(\times\) 	&1.15\(\times\) 	&1.04\(\times\) 	&1.00\(\times\)    \\
                           & After 2nd              &1.33\(\times\) 	&1.16\(\times\) 	&1.06\(\times\) 	&1.02\(\times\)                           \\ \hline
    \multirow{2}{*}{32}    & 1st                    &1.17\(\times\) 	&1.16\(\times\) 	&1.04\(\times\) 	&1.01\(\times\)    \\
                           & After 2nd              &1.37\(\times\) 	&1.12\(\times\) 	&1.04\(\times\) 	&1.02\(\times\)                           \\ \hline
    \multirow{2}{*}{64}    & 1st                    &1.19\(\times\) 	&1.16\(\times\) 	&1.04\(\times\) 	&1.00\(\times\)    \\
                           & After 2nd              &1.26\(\times\) 	&1.08\(\times\) 	&1.03\(\times\) 	&1.01\(\times\)                            \\ \hline
    \multirow{2}{*}{128}   & 1st                    &1.20\(\times\) 	&1.14\(\times\) 	&1.05\(\times\) 	&1.02\(\times\)    \\
                           & After 2nd              &1.21\(\times\) 	&1.07\(\times\) 	&1.03\(\times\) 	&1.02\(\times\)                           \\ \hline
    \multirow{2}{*}{256}   & 1st                    &1.19\(\times\) 	&1.15\(\times\) 	&1.05\(\times\) 	&1.00\(\times\)     \\
                           & After 2nd              &1.22\(\times\) 	&1.06\(\times\) 	&1.05\(\times\) 	&1.02\(\times\)                           \\ \hline
    \multirow{2}{*}{512}   & 1st                    &1.20\(\times\) 	&1.13\(\times\) 	&1.00\(\times\) 	&1.00\(\times\)    \\
                           & After 2nd              &1.21\(\times\) 	&1.09\(\times\) 	&1.04\(\times\) 	&1.14\(\times\)                           \\ \hline
    \multirow{2}{*}{1024}  & 1st                    &1.19\(\times\) 	&1.24\(\times\) 	&1.41\(\times\)     & N/A           \\
                           & After 2nd              &1.28\(\times\) 	&1.13\(\times\) 	&1.05\(\times\)     & N/A           \\
    \bottomrule
    \end{tabular}
\end{subtable}
\end{table}
\end{landscape}
%
%
As described in section \ref{section:dpt}, DPT is based on a grid search algorithm to find the optimal values of parameters for the dataloader.
The CIFAR-10 dataset is used to initially evaluate the 
 performance of DPT.
The DPT was executed for the range of workers from 1 to 48 and the  range of prefetch factors from 1 to 48 using 32 batch size (usually used when using CIFAR-10).
Figure \ref{fig:grid-worker} is the plot of the normalized transfer time by prefetch factor for the number of worker, and Figure \ref{fig:grid-prefetch} shows the fluctuation when the prefetch factor is changed for the optimal number of worker which was 10. The normalization is done by dividing all transfer times by the longest transfer time for each prefetch factor.
The blue horizontal line in Figure \ref{fig:grid-worker} is the transfer time using PyTorch's default value of 6 and 2 for the number of workers and prefetch factor parameters, respectively.

Figure \ref{fig:grid-worker} shows that the optimal number of worker can lead to 1.3\(\times\) increase in performance compared using the PyTorch's default values for the dataloader.
%
The reason for the optimal number of 10 workers may be that the Pytorch main process is running on one logical core, and dataloader's main process is residing on another logical core,  which means that there are 10 free logical cores to used by 10 workers.
As seen in Figure \ref{fig:grid-prefetch}, although the fluctuation of the transfer time depending on prefetch factor is not large, the optimal value is still unpredictable. Therefore, prefetch factor will be needed to determined carefully for different environments.

\subsection{COCO Dataset}
COCO dataset is used to evaluate the impact of dataset size using various image resolutions.
The COCO dataset's resolution is resized to 80\(\times\)80, 160\(\times\)160, 320\(\times\)320, and 640\(\times\)640 for evaluation.
Table \ref{tab:optimal-worker} and \ref{tab:transfer_time} show the results for optimal number of workers and transfer time, respectively.

%
From Table \ref{tab:optimal-worker}, we can find that if dataset resolutions are low, spawning worker to CPU's full capability derives best performance.
However, as dataset size grow over 320\(\times\)320 (7.4GB), optimal worker for first epoch and from second onward tend to differ.
On the first epoch, dataloader fetches dataset from storage to main memory.
From second epoch onward, dataloader can find data set in the main memory, which leads to lower memory operation time.
However, this holds only when dataset can reside in the main memory.
When data item size becomes larger, such as the data item size becoming 640\(\times\)640 in this experiment, it seems that the number of workers becomes limited by the memory size.  Thus, the optimal number of workers is reduced for cases of second epoch onward. In the 1024 batch case for 640\(\times\)640 resolution dataset, the GPU memory could not support the data size and could not be executed.

\section{Conclusion}
\label{section:conclusion}
In this paper, Dataloader Parameter Tuner (DPT) is proposed, which is an automated framework that will determine the optimal values for number of workers created by the dataloader and prefetch factor. The optimal values are used for faster data transfer for deep learning models.
DPT initially takes machine and dataset-dependent parameters into account.
Therefore, parameters deduced by DPT can be used for datasets with similar characteristics. For executing deep learning models on larger systems such as Cloud computing platforms, it may be more important to determine these optimal values as there will be a lot more CPUs and GPUs.

\bibliographystyle{ACM-Reference-Format}
\bibliography{references.bib}

\end{document}